\documentclass[a4paper,11pt]{article}
\pdfoutput=1
\usepackage{jinstpub}

\usepackage{lineno}

\usepackage[normalem]{ulem}
\usepackage{color}
\definecolor{blue}{rgb}{0,0,1}
\definecolor{red}{rgb}{1,0,0}

\begin{document}

\title{\boldmath Novel method for in-situ drift velocity measurement in 
  large volume TPCs: the Geometry Reference Chamber of the 
  NA61/SHINE experiment at CERN}

\author[a,*]{Andr\'as L\'aszl\'o,}
\author[a,b]{\'Ad\'am Gera,}
\author[a]{Gerg\H{o} Hamar,}
\author[a,c]{Botond P\'alfi,}
\author[d]{Piotr Podlaski,}
\author[e,f]{Brant Rumberger,}
\author[a]{and Dezs\H{o} Varga}

\affiliation[a]{HUN-REN Wigner Research Centre for Physics, Budapest, Hungary}
\affiliation[b]{Current affiliation: GSI, Darmstadt, Germany}
\affiliation[c]{E\"otv\"os University, Budapest, Hungary}
\affiliation[d]{University of Warsaw, Warsaw, Poland}
\affiliation[e]{University of Colorado Boulder, Boulder, CO, USA}
\affiliation[f]{Current affiliation: Trimble Inc., Westminster, CO, USA}
\affiliation[*]{Corresponding author.}

\emailAdd{
  laszlo.andras@wigner.hun-ren.hu,
  gera.adam@wigner.hun-ren.hu,
  hamar.gergo@wigner.hun-ren.hu,
  palfi.botond@gmail.com,
  piotr.podlaski@cern.ch,
  brant.rumberger@gmail.com,
  varga.dezso@wigner.hun-ren.hu
}

\abstract{
  This paper presents a novel method for low maintenance, low ambiguity 
  in-situ drift velocity 
  monitoring in large volume Time Projection Chambers (TPCs). The method 
  was developed and deployed for the $40\,\mathrm{m}^{3}$ TPC tracker system 
  of the NA61/SHINE experiment at CERN, which has a one meter of drift length. 
  The method relies on a low-cost multi-wire proportional chamber placed 
  next to the TPC to be monitored, downstream with respect to the particle flux. 
  Reconstructed tracks in the TPC are matched to hits in the monitoring 
  chamber, called the Geometry Reference Chamber (GRC). Relative differences 
  in positions of hits in the GRC are used to estimate the drift velocity, 
  removing the need for an accurate alignment of the TPC to the GRC. 
  An important design requirement on the GRC was minimal added complexity 
  to the existing system, 
  in particular, compatibility with Front-End Electronics 
  cards already used to read out the TPCs. Moreover, the GRC system was 
  designed to operate both in large and small particle fluxes. The system is 
  capable of monitoring the evolution of the drift velocity inside the TPC 
  down to a one permil precision, with a few minutes of data collection.
}

\keywords{Time projection chambers (TPC),
  Particle tracking detectors (Gaseous detectors),
  Large detector systems for particle and astroparticle physics}

\arxivnumber{2405.01285}

\maketitle
\flushbottom

\section{Introduction}
\label{sec:introduction}

The NA61 experiment, also called the SPS Heavy Ion and Neutrino Experiment (SHINE) 
\cite{abgrall2014}, is a large-acceptance hadron spectrometer experiment at 
the Super Proton Synchrotron (SPS) accelerator at CERN. 
Its physics program includes measurements for heavy-ion physics, as well as 
measurements on particle production in hadron-nucleus collisions, 
with an emphasis on the application of those measurements in flux predictions 
for long-baseline neutrino oscillation experiments 
(T2K, DUNE, NOvA, Hyper-Kamiokande) \cite{abe2011,abi2020,ayres2007,abe2018} and 
for cosmic ray observatories (such as Pierre Auger Observatory) \cite{aab2015}. 
These physics programs demand the operation of 
NA61/SHINE in both high and low particle multiplicity environments.

The main components of the NA61/SHINE experiment are two large superconducting 
bending magnets and a system of large volume 
Time Projection Chambers (TPCs). As the drift dimension 
of the TPCs is considerably large (${\gtrsim}1\,$meter), 
in order to achieve necessary 
position resolution in the drift direction, the drift velocity 
inside the TPCs needs to be monitored at the permil level. 
Since the drift velocity slowly evolves with time, mostly due to variations 
in the ambient conditions or gas composition, 
its evolution needs to be followed every few minutes of data taking. 
In this paper, we describe a cost-efficient supplementary detector system 
specifically developed for this purpose.

One of the default methods for monitoring drift velocity in a TPC 
is to analyze its exhaust gas with a small probe chamber 
\cite{kuich2019,abgrall2014,afanasiev1999,cattai1989} 
and then predict the drift velocity inside the TPC via the 
known pressure, temperature and drift field in the TPC gas. 
This method dates back to the large-volume TPCs of ALEPH and DELPHI at LEP \cite{atwood1991,brand1989}. 
NA61/SHINE also uses this technique to monitor the drift velocity in its TPCs, 
but due to the limited absolute 
systematic accuracy of the method, it is mainly used for monitoring the 
time stability of the gas composition.\footnote{The so-called normalized 
drift velocity, estimated via exhaust analyzer, is the hypothetical 
drift velocity extrapolated to a fixed pressure, fixed temperature, 
and to a fixed drift field. That quantity, although is not directly related 
to the actual drift velocity inside the TPC, but is rather useful for monitoring 
the time stability of the working gas composition.} 
The exhaust method is prone to uncertainties in pressure and temperature 
measurements in the large-volume chambers, and is also vulnerable to 
contaminations along the exhaust collection and system outgassing.

An other typical method applied in a number of experiments is the so-called 
bottom point method: the drift time of the last points of 
tagged cathode-exiting tracks can be used for drift velocity estimation, 
provided that the delay of the trigger signal to the TPC readout is known 
to a good accuracy, along with a precise estimate for the effective drift 
length. In NA61/SHINE we do not rely on this method, primarily because it is 
prone to systematic uncertainties of the effective drift length, but more 
importantly of the trigger delay, which is not even known a priori. 
A further method, often applied in large volume TPC experiments, relies 
on imaging start and end points of cosmic ray tracks. In NA61/SHINE, such 
method is not suitable, as the orientation of the TPC acceptance and drift 
direction is suboptimal with respect to the celestial directions, leading to 
large uncertainties in the cosmic track start and end point reconstruction.

Due to the systematic uncertainties of the above approaches, a 
more advanced absolute in-situ drift velocity measurement method is needed. 
For such purpose, typically a UV laser system is used: ionizing laser beam tracks 
are projected into the sensitive volume at known locations, and the drift 
velocity can then be measured from the apparent positions of these beams in 
terms of TPC drift time. 
Direct measurement methods using laser systems are common 
\cite{alme2010,afanasiev1999} and very accurate. UV laser-based 
drift velocity and geometry monitoring systems, however, are rather 
hard to implement or upgrade on existing chambers retrospectively. 
Moreover, the cost and development time for such systems is substantial. 
NA61/SHINE is not equipped with such a UV laser based system.

\begin{figure}[!ht]
  \begin{center}
    \includegraphics[height=6cm]{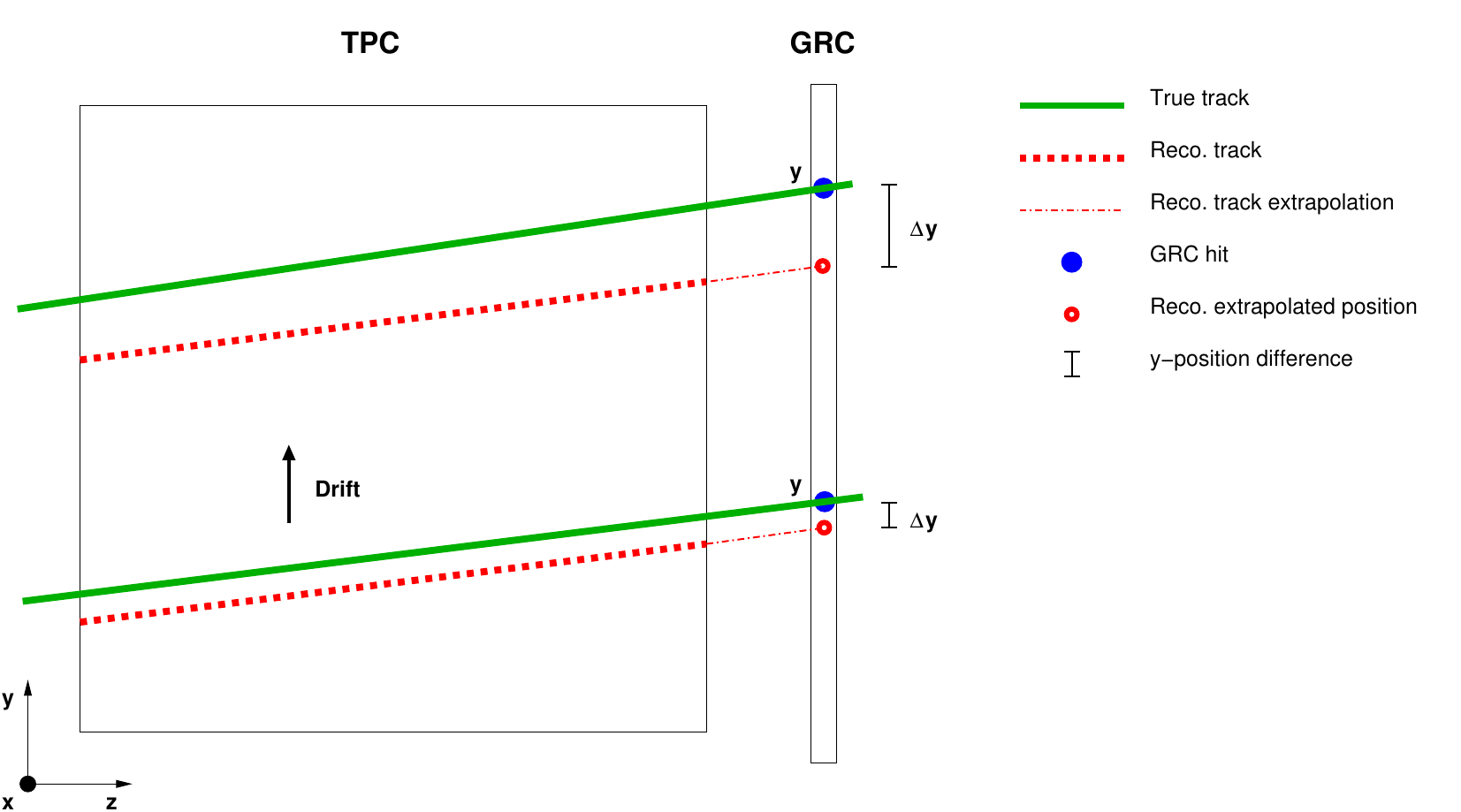}
  \end{center}
  \caption{(Color online) Illustration of the Geometry Reference Chamber (GRC) 
    concept. The particle tracks in the TPC chamber are reconstructed with 
    an initial estimate for the drift velocity and are paired to the hits 
    of a segmented detector, the GRC. In case the initial approximate drift velocity 
    estimate differs from the true drift velocity inside the TPC, the 
    mismatch along the drift coordinate gets gradually worse at increasing 
    drift depths. The slope of the drift coordinate mismatch against the 
    drift coordinate as measured by the GRC provides the estimate for the corresponding 
    multiplicative correction factor. 
    (The figure shows the projection orthogonal to the magnetic bending plane, 
    in case the TPC is used in a magnetic spectrometer experiment, such as the NA61/SHINE.)
  }
  \label{fig:grcconcept}
\end{figure}

Along with other CERN experiments, the NA61/SHINE facility was also 
significantly upgraded during the Long Shutdown 2 (LS2) period (2019-2021) of 
the accelerator complex. Since during the LS2 the 
subdetectors were moved, 
precise alignment calibration of the chambers was required, 
as was a system for accurately and reliably determining their drift velocities. 
As the TPCs were already constructed, minimally-invasive measurement systems 
were preferred. Our choice was a simple solution: a planar detector 
with fixed segmentation along the TPC drift direction, 
called the Geometry Reference Chamber (GRC), was to be placed downstream 
of the TPC system with respect to the particle flux. 
TPC tracks reconstructed with an approximately known 
drift velocity are extrapolated to the GRC and paired with the GRC hits. 
If the drift velocity estimate differs from the true drift velocity 
inside the TPC, the mismatch of the drift coordinate of the hits becomes 
worse with increasing drift depth. The ratio between the initial approximate 
drift velocity and the actual drift velocity in the chamber can then be estimated from 
the slope of this mismatch scatter plot. 
That is, the GRC is used as a differential length scale in order to 
convert the drift time, measured by the TPC readout, to drift depth. 
It is quite advantageous that the GRC 
method is differential: an accurate alignment of the TPC against the GRC is 
not needed. The concept is illustrated 
in Fig.\ref{fig:grcconcept}.\footnote{Once the drift velocity has been 
calibrated with such method in one of the TPC chambers in the detector complex, 
the other chambers can be calibrated against each-other, using the 
already calibrated chambers as geometry reference. Moreover, the remaining 
geometry calibration constants of the TPC system can also be estimated, 
such as the trigger delay ($t_{0}$), alignment shift and alignment 
angles for each chamber, see Appendix~\ref{sec:appT0Y0}~and~\ref{sec:appAlign}.}

In order to implement the GRC concept, several design requirements were considered. 
To reduce the development time and cost of the readout electronics, 
a solution using the existing TPC readout Front-End Electronics (FEE) was 
chosen. 
To reduce the number of readout channels, we decided to implement 
a simple planar Multi-Wire Proportional Chamber (MWPC) 
with Cartesian readout. This setting required two issues to be addressed. 
Firstly, the signal formation in the GRC had to be slowed down 
sufficiently such that the signal persists until the TPC electronics begins 
to sample the charges. This is not entirely trivial to achieve, since 
TPC readout electronics typically allow for a microsecond-level trigger delay 
with respect to the particle passage time, 
during which the signal may already disappear from the GRC. 
In order to mitigate this issue, a drift volume with adjustable drift field 
was introduced in the GRC design, transverse to the particle passage direction, 
and with that the GRC signal duration could be controlled. 
Secondly, the Cartesian readout scheme tends to be problematic whenever the particle 
occupancy is large. In order to overcome this limitation, the GRC acceptance 
was designed to be variable: for low-multiplicity collisions, the 
entire GRC acceptance can be used, whereas for high-multiplicity collisions, 
individual anode wires (sense wires) of the GRCs can be disabled, down to a single amplifying 
wire. This helps to reduce the number of combinatorial ghost hits.

The structure of the paper is as follows. 
In Section~\ref{sec:NA61ExperimentalFacility} an overview of the NA61/SHINE experimental facility is given. 
In Section~\ref{sec:grc} the GRC design requirements, parameters, and implementation are discussed. 
In Section~\ref{sec:performance} the operational performance is shown. 
In Section~\ref{sec:conclusion} a conclusion is provided. 
Appendices~\ref{sec:appT0Y0} and \ref{sec:appAlign} discuss some details of systematic 
errors and the estimation of further geometry-related TPC calibration constants, 
where the GRC chamber plays a role.

\section{The NA61/SHINE experimental facility}
\label{sec:NA61ExperimentalFacility}

The NA61/SHINE \cite{abgrall2014} is a fixed-target hadron spectrometer 
experiment at the CERN SPS accelerator. 
Large parts of its tracking devices were inherited 
from a previous experiment called NA49 \cite{afanasiev1999}. Its 
physics program covers the study of strongly-interacting 
matter via heavy-ion collisions, and 
measurements of identified particle production spectra 
in hadron-nucleus collisions as reference data for flux prediction 
in long baseline neutrino experiments and large area cosmic ray 
observatories.

\begin{figure}[!ht]
  \begin{center}
    \includegraphics[width=14cm]{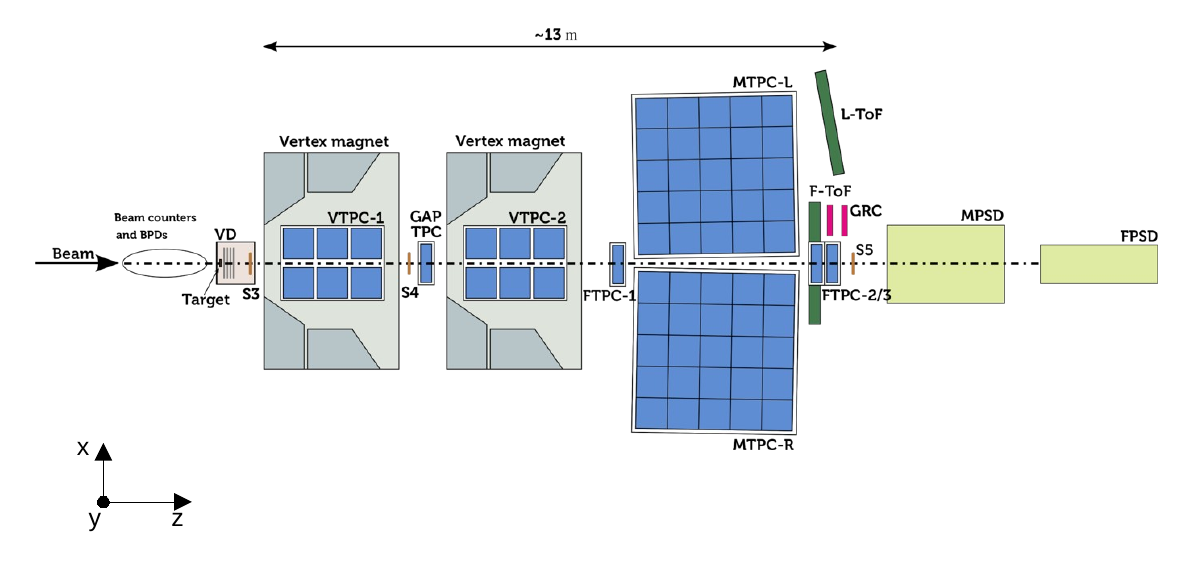}
  \end{center}
  \caption{(Color online) The NA61/SHINE detector configuration,
    after the Long Shutdown 2 (LS2) upgrade, i.e.\ since 2022. 
    The drawing shows the top view of the bending plane. That is, the 
    approximately homogeneous magnetic field inside the Vertex-I and 
    Vertex-II magnets is perpendicular to the figure (${\sim}y$ axis). The 
    electric drift field of the TPCs is also perpendicular to the figure (${\sim}y$ axis). 
    The GRC system for in-situ drift velocity monitoring was also 
    conceptualized and introduced during the LS2. The GRC system, as of now, 
    consists of two adjacent identical chambers (GRC-1 and GRC-2), for 
    redundancy reasons. As the figure is a top view of the experiment, 
    the shorter dimension of the GRCs is seen in this outline.
  }
  \label{fig:DetectorSetup}
\end{figure}

An overview of the NA61/SHINE experiment is shown in Fig.\ref{fig:DetectorSetup}. 
Two large superconducting bending magnets (Vertex-I and II) are responsible for 
particle deflection for charge and momentum determination. Their total maximum bending 
power is ${\sim}9\,\mathrm{Tm}$ (up to $1.5\,\mathrm{Tesla}$ magnetic field 
in Vertex-I and $1.1\,\mathrm{Tesla}$ in Vertex-II). 
A target holder with target-in / target-out moving capability sits just 
upstream of the first Vertex TPC, upstream/downstream always 
meant with respect to the beam direction throughout the paper. 
Thin targets can be placed inside a silicon Vertex Detector (VD) for precise vertex determination. 
NA61/SHINE also has the ability to measure interactions in extended 
replica targets for long baseline neutrino experiments. The tracking devices for spectrometry 
are composed of eight large volume TPCs (total ${\sim}40\,\mathrm{m}^{3}$ and 
${\sim}1\,\mathrm{m}$ drift length), 
capable of performing both tracking and $\mathrm{d}E/\mathrm{d}x$ measurements. 
A Multigap Resistive Plate Chamber (MRPC) based Time-of-Flight wall (L-ToF) provides further particle identification 
(PID) capabilities around mid-rapidity, whereas a scintillator based 
Time-of-Flight wall (F-ToF) covers the forward phase space, 
enabling two-dimensional (ToF+$\mathrm{d}E/\mathrm{d}x$) separation of 
particle species at large parts of the acceptance. 
A calorimeter is placed at the end of the beamline, called the Projectile 
Spectator Detector (PSD), which helps to characterize collision centrality in 
heavy-ion collisions, and consists of two compartments (Main- and Forward-PSD, 
also called MPSD and FPSD) for optimal shower containment. 
Upstream of the target position, a set of beam position detectors (BPDs) 
provide beam tracking information, while 
scintillator and Cherenkov detectors serve as beam trigger with PID 
capability (not shown in the figure). 
On the beamline between VTPC-1 and the GapTPC, a small plastic scintillator with a 
$1\,\mathrm{cm}$ diameter serves as 
an interaction trigger in most collision types (S4). In rare run settings, a 
very similar scintillator (S5) just upstream of MPSD is used instead. 
For heavy-ion runs, the scintillator (S3) just downstream of VD is used for 
interaction definition through detecting loss of charge 
(i.e., by sensing a decreased $Z^{2}$ with respect to the beam nuclei). 
The VD, the new silicon BPDs, the MRPC-based L-ToF, 
the MPSD and FPSD were introduced during the LS2 upgrade 
period between 2019-2021. Also during LS2, the data acquisition (DAQ) was 
upgraded, allowing for an event recording rate up to 
$1.8\,\mathrm{kHz}$ for the detector complex.

The concept of the Geometry Reference Chambers (GRCs), 
just downstream of MTPC-L, was also developed 
during the LS2 upgrade period. Its lightweight design was motivated by 
short development time, and relatively fast and low risk implementability. 
In addition, the TPCs were 
moved during the LS2 upgrade, and therefore seven calibration constants 
apart from the drift velocity must be re-determined for each TPC. 
These are the trigger delay, the three-dimensional alignment shifts, 
and the three alignment angles, per chamber, independently. The large 
number of unknown calibration constants are hard to determine 
without an absolute reference. The GRC enables unambiguous 
determination of all seven parameters and the drift velocity, for each 
component of the TPC system.

\section{The GRC system}
\label{sec:grc}

The NA61/SHINE TPCs have a drift length of ${\sim}1\,\mathrm{m}$ and are 
capable of resolving particle cluster positions at the ${\sim}1\,\mathrm{mm}$ 
at the extremum of the drift volume. Therefore, in order to reduce the 
systematic errors of the position measurements below the position resolution, 
the drift velocity has to be determined with permil accuracy. 
The drift velocity in a typical TPC system is affected by the working 
gas composition, the electric drift field applied to the TPC, the gas temperature, 
and the gas pressure. 
Most large TPC systems, such as the TPCs in NA61/SHINE experiment 
or the one in ALICE \cite{alme2010} at CERN's LHC, or in STAR \cite{abele2003,anderson2003} at BNL's RHIC, 
are equilibrated 
with the atmospheric pressure, with a very slight constant overpressure applied in 
order to avoid air infiltration. Since the gas composition, drift field and 
the temperature is typically stabilized, the shortest time scale variations 
are caused by meteorological changes in the ambient air pressure. This means 
that the time scale of anticipated substantial drift velocity variations are 
not shorter than about $5$ minutes. 
Therefore, the design goal of the GRC system was to 
have a drift velocity measurement inside MTPC-L with a permil 
statistical error for about every $5$ minute time window 
during data taking. Given the worst-case minimal particle flux, 
this requirement gives the lower bound to the acceptance (area) of the GRC 
in case of low-multiplicity collision types. 
In order to minimize the necessary area of the GRC, a location was chosen 
with the maximum possible particle flux passing through MTPC-L, but avoiding 
the vicinity of the beam spot. This justifies the chosen placement of the GRC 
system, shown in Fig.\ref{fig:DetectorSetup}. At the selected location, 
the worst-case minimum particle flux was estimated using formerly recorded 
proton-proton collision data at $13\,\mathrm{GeV}/c$ beam momentum. 
With about a factor of two safety margin, this resulted in a 
$40\,\mathrm{cm}$ $\times$ $120\,\mathrm{cm}$ sensitive area, with the longer 
dimension fully covering the drift direction of MTPC-L. For redundancy 
reasons, two identical GRC stations (GRC-1 and GRC-2) were placed in an 
adjacent position (see Fig.\ref{fig:DetectorSetup}).

\begin{figure}[!ht]
  \begin{center}
    \includegraphics[height=7cm]{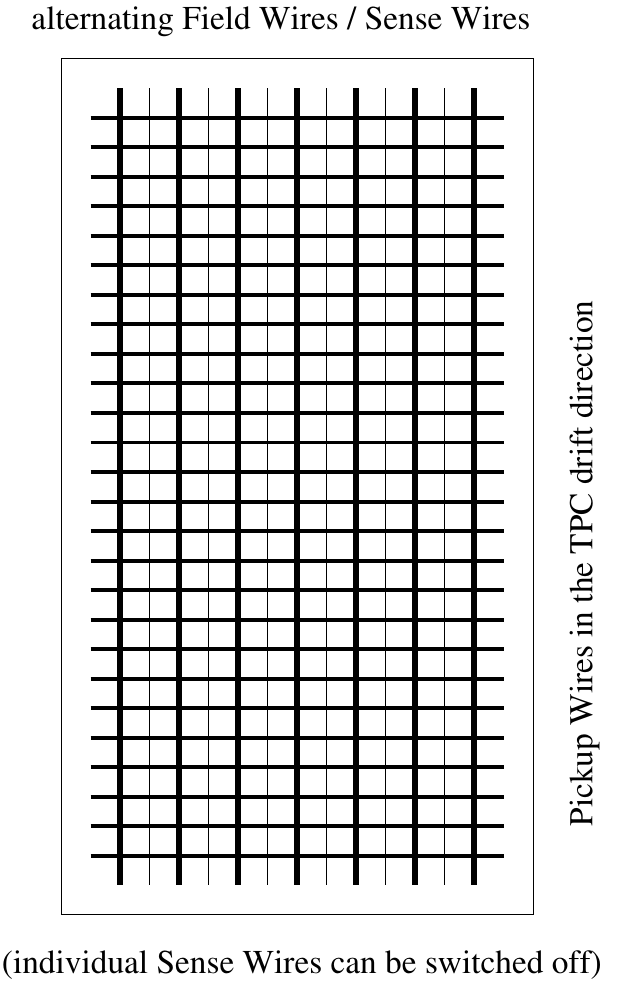}
    \hspace*{2cm}
    \includegraphics[height=7cm]{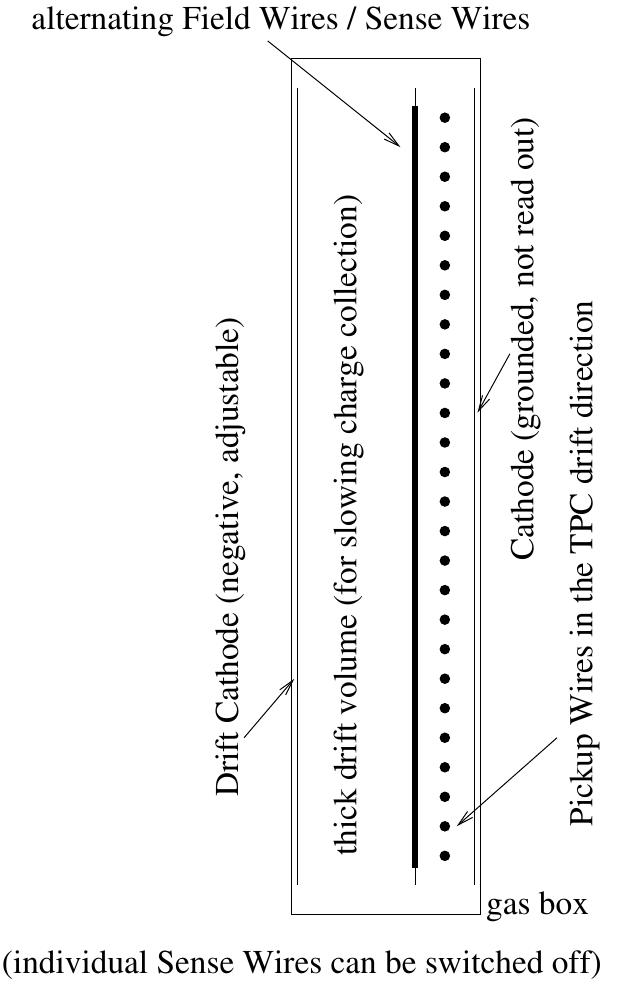}
  \end{center}
  \caption{Left panel: sketch of a GRC chamber as seen from 
    the direction of the arriving particle flux (not to scale). The transverse direction is 
    covered by the field wire / sense wire plane, where the electron multiplication 
    takes place on the sense wires, and their mirror charges on the field wires 
    are read out. The TPC drift direction is covered by the pickup wire plane, 
    also seeing the mirror charges of the charge clouds on the sense wires. 
    This provides the other Cartesian coordinate for readout. 
    Individual sense wires can be connected to high voltage or ground using 
    configurable high-voltage jumpers (not drawn on the figure). 
    Right panel: sketch of a 
    GRC as seen from side view (not to scale). The thickness of the drift region 
    is larger in the direction of the particle flux, and a drift cathode 
    electrode with adjustable potential is used. This combination makes sure 
    that the charge collection can be sufficiently slowed down such that 
    the signal in the chamber persists until the TPC readout electronics, used for the 
    GRC readout, begins to measure.
  }
  \label{fig:grcsketch}
\end{figure}

For cost efficiency reasons, the GRC implementation was chosen to be a robust 
MWPC with Cartesian readout loosely based on the design \cite{varga2016,varga2022}, 
which was originally optimized for cosmic muon imaging purposes. 
Gas amplification is achieved on the anode wires (sense wires), and the transverse 
position information is read from adjacent wires (field wires) parallel to these in the same plane. 
For the vertical position information relevant for drift velocity monitoring, 
the segmentation is achieved using wires with the same functionality as cathode strips, 
perpendicular to the sense wire / field wire direction (those are called pickup wires). The design choice 
to use such wires instead of strips is due to easier construction of the 
whole chamber. 
For low-multiplicity data taking campaigns, the entire GRC acceptance is 
read out as is. For high-multiplicity runs, such as 
heavy-ion data taking periods, the Cartesian combinatorial 
background becomes an issue. In order to mitigate this, the design allows 
for high-voltage jumpers used to turn off individual sense wires: 
the sense wires can be connected either to their operational voltage 
or to the chamber ground. 
The jumpers are configured manually at the start of a data taking campaign. 
Turning off sense wires has the effect of narrowing the transverse acceptance, 
eventually down to a single active sense wire. 
This feature completely eliminates the Cartesian combinatorics, 
preserving the drift velocity determination accuracy in high multiplicity 
runs. The concept is sketched in the left panel of Fig.\ref{fig:grcsketch}.

In order to reduce cost and development time for the readout electronics, 
the GRCs were designed to be read out using the existing TPC FEEs. 
Two TPC FEE cards are enough to read out the induced charges on the 
$32$ field wire channels $\times$ $192$ pickup wire channels in each GRC. A natural issue 
with this solution is that the TPC, being a relatively slow device, allows 
for a relatively large trigger delay. Typical TPC trigger and readout electronics are 
only optimized for making meaningful charge sampling measurements 
$1-2\,\mu\mathrm{s}$ after the beam passage. This relatively large but constant 
time delay is due to trigger logic decision time, signal arrival time on the 
signal cables, as well as the startup time of the FEE logic. 
In order to make sure that the signal in the GRC persists until the TPC 
FEE begins to take meaningful charge samples, a $30\,\mathrm{mm}$ thick 
drift volume was added.  This has the effect of extending charge collection over time. 
In order to control the signal collection timescale, an adjustable drift field electrode was added 
to the design. The signal stretching concept is shown schematically 
in the right panel of Fig.\ref{fig:grcsketch}.\footnote{Due 
to the mentioned thickened charge collection volume, the charge collection 
direction slightly differs from the the particle incidence direction in 
case of tracks arriving non-orthogonally at the GRC surface. 
This causes a small angle-dependent position bias, 
which can be easily corrected for, using the incidence angles estimated by the TPC track piece. 
Due to the NA61/SHINE geometry, the pertinent bias is of the order of 
${\sim}1\,\mathrm{mm}$ at the extremes, and vanishes 
toward the middle of the GRC acceptance. Moreover, in the NA61/SHINE geometry, 
this bias is second order, therefore it happens to cancel in the GRC-based 
drift velocity calibration procedure.}
The GRC front-end readout system is fully integrated with the NA61/SHINE 
Data Acquisition System (DAQ). GRC-specific event reconstruction software 
is fully integrated with the NA61/SHINE offline software (ShineOffilne) \cite{sipos2012}. 
The fundamental parameters of the MWPC design of the GRCs are listed 
in Table~\ref{table:grcparams}. 

\begin{table}[!ht]
  \begin{center}
    \begin{tabular}{r|l}
     chassis                                      & FR4 bars and typical PCB board material \\
     typical working gas                          & $\mathrm{Ar}(80):\mathrm{CO}_{2}(20)$ at $5\,\mathrm{l/h}$ flow \\
     typical drift cathode potential              & $0\,\mathrm{V}$ \\
     field wire and pickup wire potential         & $0\,\mathrm{V}$ \\
     typical sense wire potential                 & $1580\,\mathrm{V}$ \\
     pickup wire                                  & CuZn40 (EDM Tec) $100\,\mu{}\mathrm{m}$, $70\,\mathrm{g}$ tension \\
     field wire                                   & CuZn40 (EDM Tec) $100\,\mu{}\mathrm{m}$, $70\,\mathrm{g}$ tension \\
     sense wire                                   & Au coated W (Luma Metall) $20\,\mu{}\mathrm{m}$, $15\,\mathrm{g}$ tens. \\
     field wire to sense wire distance            & $6\,\mathrm{mm}$ \\
     distance between adjacent pickup wires       & $3\,\mathrm{mm}$ (read out pairwise) \\
     FW/SW to pickup wire plane distance          & $10\,\mathrm{mm}$ \\
     pickup wire plane to cathode backplane dist. & $2\,\mathrm{mm}$ \\
     drift cathode to FW/SW plane                 & $30\,\mathrm{mm}$ \\
     number of sense wires                        & $31$ \\
     number of field wires                        & $32$ \\
     number of pickup wire pairs                  & $192$ \\
    \end{tabular}
  \end{center}
  \caption{Design parameters of the GRC MWPCs.}
  \label{table:grcparams}
\end{table}

\begin{figure}[!ht]
  \begin{center}
    \includegraphics[height=9cm]{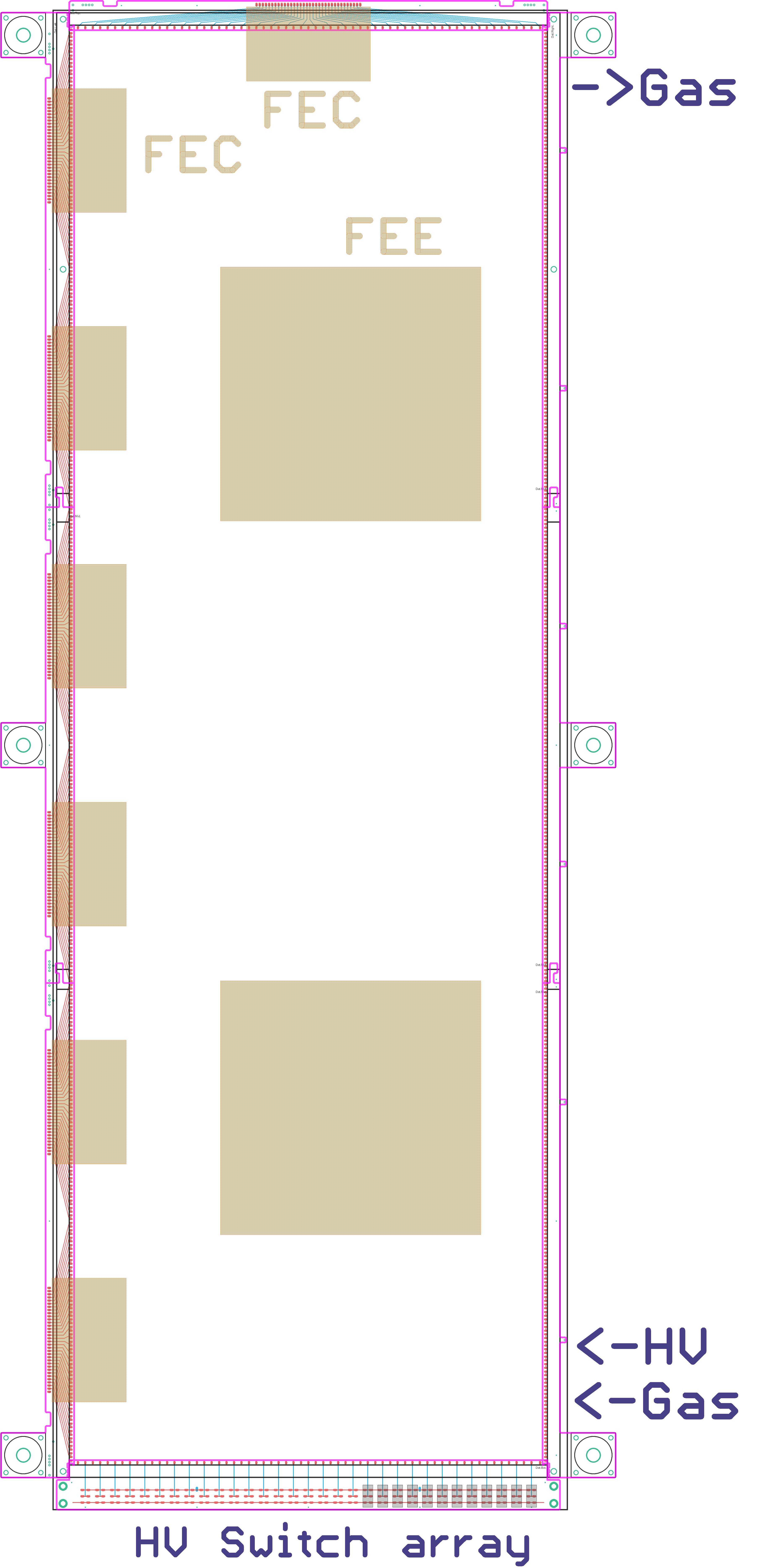}
    \hspace*{2cm}
    \includegraphics[height=9cm]{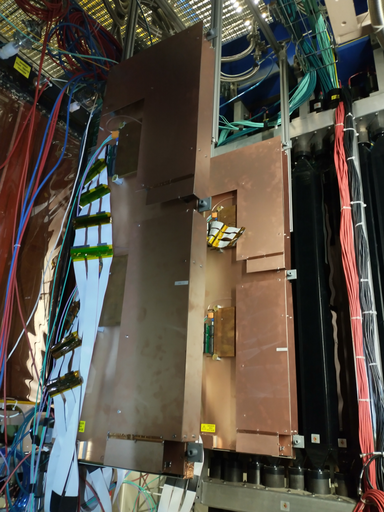}
  \end{center}
  \caption{(Color online) Left panel: the PCB CAD snapshot of the sensor layer 
    of a GRC chamber. The six holders for the optical geodesic survey 
    targets are well visible. By design, these are precisely 
    relatable to the internal wire geometry. 
    Right panel: the photograph of the installed GRC system as seen 
    looking upstream. The black wall is the F-TOF, covering the downstream 
    plane of MTPC-L. The upstream chamber is GRC-1, whereas the downstream 
    chamber is GRC-2. 
  }
  \label{fig:grcPhoto}
\end{figure}

Since the GRC performs a differential measurement along the TPC drift direction, 
precise alignment of the GRCs 
with respect to the TPC is not critical. 
However, if the TPC-GRC alignment is known, the GRC system can be used to 
determine a further calibration constant: the trigger delay ($t_0$). 
In order to make use of this possibility, care was taken in the design that the 
position of the optical survey target for geodesic localization of the GRCs 
can be accurately related to the internal wire positions. A CAD sketch of 
the sensor plane and a photograph of the installed GRC system is seen 
in Fig.\ref{fig:grcPhoto}.

\section{Drift velocity calibration using the GRC}
\label{sec:performance}

Drift velocity calibration via the GRC is based on 
track matching between MTPC-L and the GRC as sketched in 
Fig.\ref{fig:grcconcept}, and subsequently using GRC as a differential length 
scale.\footnote{For combinatorial background suppressing reasons, 
in practice global tracks are used, 
which are associated to a collision in the target (global main-vertex tracks). 
These can be already reconstructed with approximate estimates for the 
calibration parameters, such as the drift velocity. Then, these global 
main-vertex tracks are dissected to local tracks, and the parameter mismatch 
of the local track pieces in the adjacent TPC chambers are evaluated for 
calibration. For MTPC-L, the mismatch against the GRC hits are used for 
calibration.}
If the drift velocity estimate differs from the true drift velocity, 
the TPC drift coordinate versus GRC coordinate mismatch has a slope as a 
function of the GRC coordinate. This slope can be used for calibration. 
Tracks are typically collected over the course of 5 minutes of data taking, 
providing a drift velocity estimate spanning those 5 minutes.

\begin{figure}[!ht]
  \begin{center}
    \includegraphics[height=5cm]{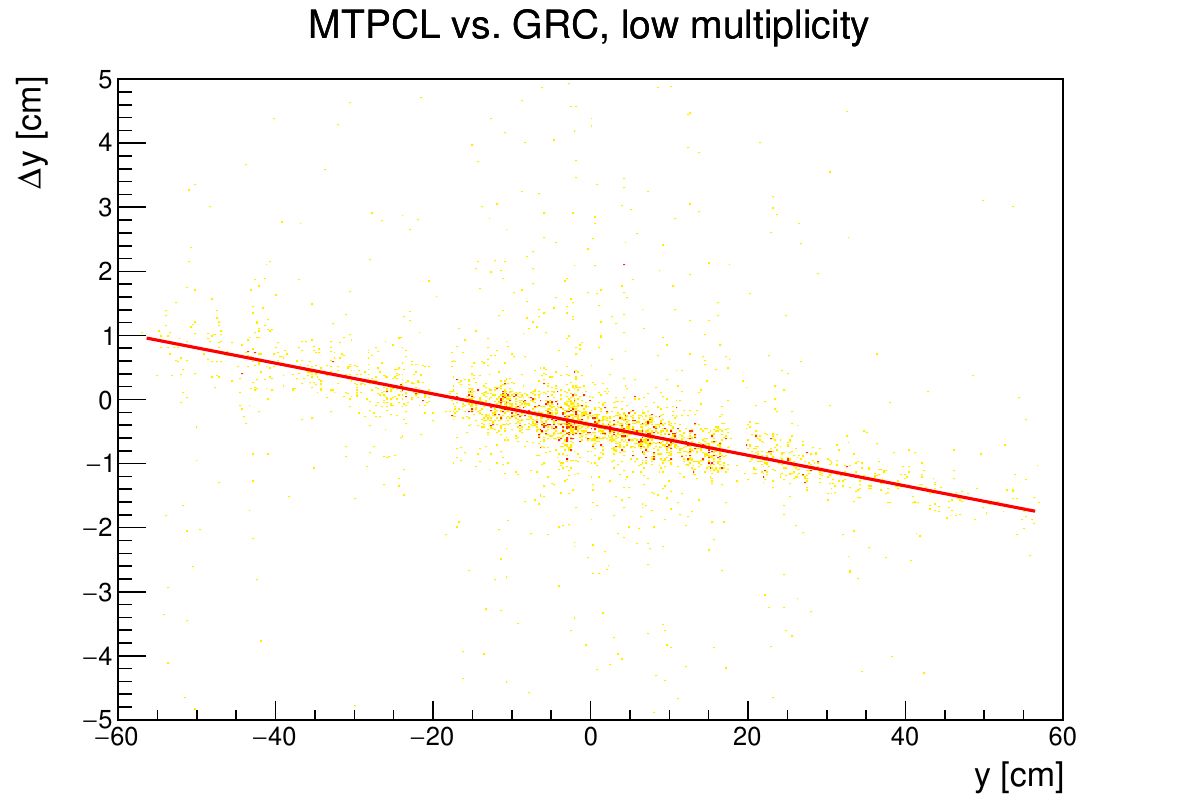}

    \vspace*{0.5cm}
    \includegraphics[height=5cm]{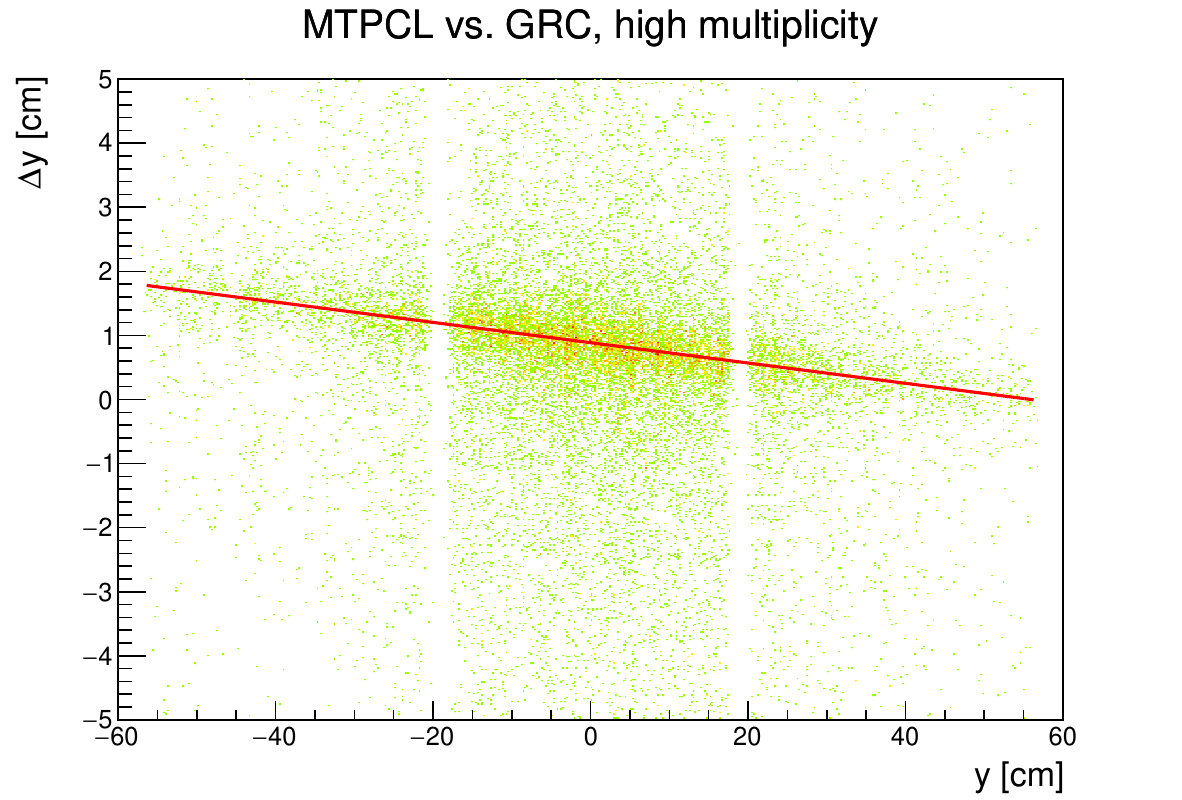}
    \includegraphics[height=5cm]{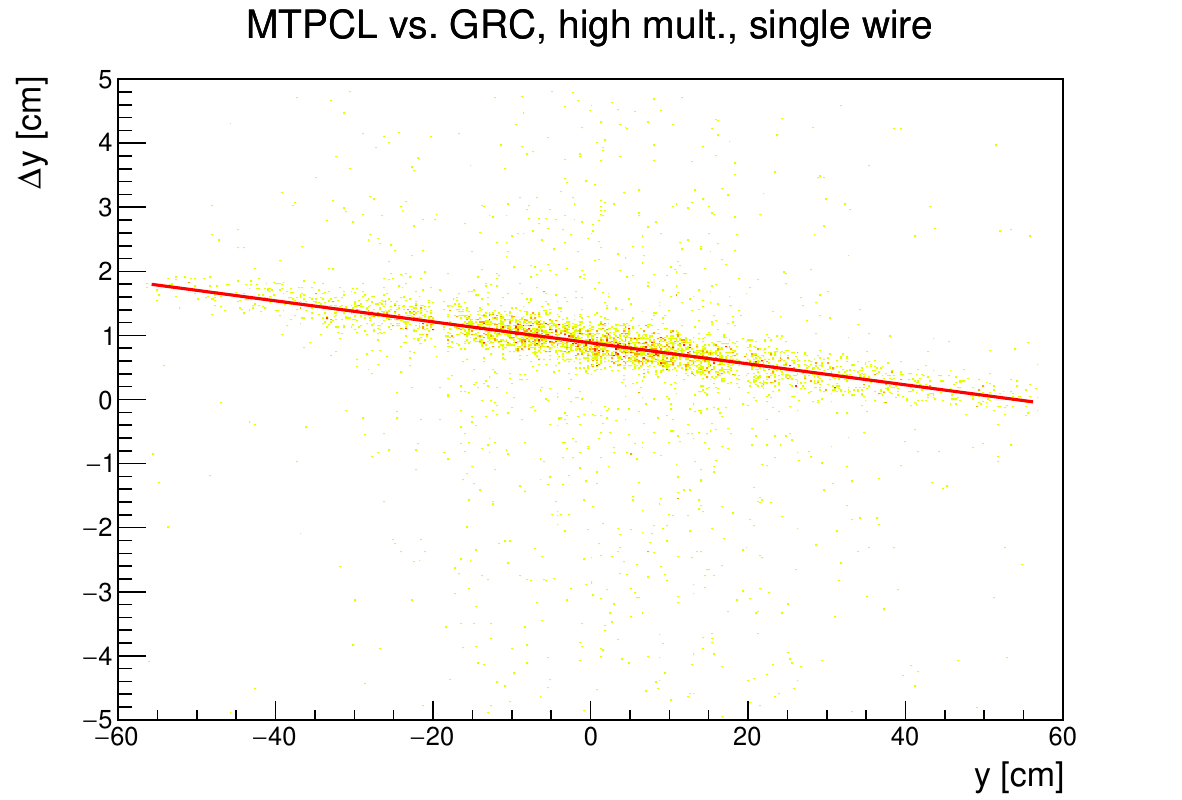}
  \end{center}
  \caption{(Color online) Top panel: an example for the 
    $\Delta{y}$ versus $y$ drift velocity calibration method on a 
    real data sample in a $5$-minute time window. 
    The corresponding data set is a low multiplicity run (proton-carbon collisions 
    at $120\,\mathrm{GeV}/c$ beam momentum, recorded in 2023). 
    Here, the full GRC acceptance was used. 
    Bottom left panel: the equivalent plot for a high multiplicity run, 
    with the full GRC acceptance (Pb+Pb collisions at 
    $150\,A\mathrm{GeV}/c$ beam momentum, recorded in 2022). 
    Bottom right panel: the same high multiplicity data set 
    with only a single sense wire used, thus eliminating the Cartesian combinatorial 
    ghost hits in GRC. Note the substantially reduced background. 
    The color scale of the scatter plot is merely to emphasize hit density for the eye. 
    The line on the plots corresponds to a straight line fit to the 
    scatter data, using the background tolerant LTS method. 
    The slopes of the fitted lines yield the drift velocity correction.
  }
  \label{fig:dYvsY}
\end{figure}

The calibration equations can be derived as follows. 
The NA61/SHINE coordinate convention, as seen in Fig.\ref{fig:DetectorSetup}, 
is that the global Cartesian $x$ coordinate is from MTPC-R to MTPC-L, the global $y$ coordinate is from down to up, 
and the global $z$ coordinate is along the nominal beamline from upstream toward 
downstream. Similarly oriented $x$, $y$, $z$ Cartesian local coordinates are 
also used within each chamber. The TPC drift coordinate is the $y$ 
local coordinate, by convention, and the drift is from negative to positive 
direction, whereas the local $x$ coordinate is along the readout pads of a 
pad row (along the sense wires of the TPC), and the local $z$ coordinate is 
stepping in between pad rows. For a TPC chamber, one has that
\begin{eqnarray}
 y_{\mathrm{nom}}  & = & y_{\mathrm{anode},\mathrm{nom}}  - (t_{0,\mathrm{nom}} +t_{\mathrm{drift,raw}})\cdot v_{\mathrm{drift,nom}}, \cr
 y_{\mathrm{true}} & = & y_{\mathrm{anode},\mathrm{true}} - (t_{0,\mathrm{true}}+t_{\mathrm{drift,raw}})\cdot v_{\mathrm{drift,true}}.
\label{eqDrift}
\end{eqnarray}
Here $y_{\mathrm{nom}}$ is the reconstructed drift coordinate assuming a 
nominal estimate during reconstruction for the position of the TPC amplification plane at 
$y_{\mathrm{anode},\mathrm{nom}}$, the anode meaning the field wire / sense 
wire plane of the TPC. The symbol $t_{0,\mathrm{nom}}$ stands for a nominal estimate for 
the trigger delay during reconstruction. $t_{\mathrm{drift,raw}}$ is the measured particle track 
cluster position in terms of drift time as measured by the TPC readout. 
$v_{\mathrm{drift,nom}}$ is the nominal estimate for the electron 
drift velocity in the TPC during reconstruction. The quantities labeled by $\,{()}_{\mathrm{true}}\,$ 
are the true but unknown values of these calibration factors. 
The negative sign in Eq.(\ref{eqDrift}) is due to the coordinate convention in NA61/SHINE. 
The coordinate $y_{\mathrm{nom}}$ measured by the TPC along the drift direction 
corresponds to a GRC measurement $y_{\mathrm{true}}$, for each track 
matched to a GRC hit, as seen in Fig.\ref{fig:grcconcept}. The matching 
is done via a typically ${\sim}3\,\mathrm{cm}$ wide tolerance window. 
By eliminating the running variable $t_{\mathrm{drift,raw}}$ in Eq.(\ref{eqDrift}), one infers
\begin{eqnarray}
 \Delta{y} & = & \underbrace{\left(v_{\mathrm{drift,nom}}/v_{\mathrm{drift,true}}-1\right)}_{\mathrm{slope}}\cdot y_{\mathrm{true}} \cr
 \Bigg. & & \qquad \;+\; \underbrace{\left(y_{\mathrm{anode,nom}}-v_{\mathrm{drift,nom}}/v_{\mathrm{drift,true}}\cdot y_{\mathrm{anode,true}}+v_{\mathrm{drift,nom}}\cdot(t_{0,\mathrm{true}}-t_{0,\mathrm{nom}})\right)}_{\mathrm{offset}}
\label{eq:dYvsY}
\end{eqnarray}
for the ensemble of TPC-GRC tracks. 
Here $\Delta{y}:=y_{\mathrm{nom}}-y_{\mathrm{true}}$ stands for the drift 
coordinate TPC-GRC mismatch and $y_{\mathrm{true}}$ 
stands for the GRC measurement. Occasionally, the label $()_{\mathrm{true}}$ 
will be suppressed where not confusing. The slope of the 
$\Delta{y}$ versus $y$ scatter plot determines the 
correction factor between the true drift velocity 
$v_{\mathrm{drift,true}}$ and the estimate $v_{\mathrm{drift,nom}}$ assumed 
during the reconstruction, thus $v_{\mathrm{drift,true}}$ can be estimated 
from the TPC-GRC mismatch data. This is shown in Fig.\ref{fig:dYvsY} 
top panel for a typical low-multiplicity data set (proton-carbon collisions 
at $120\,\mathrm{GeV}/c$ beam momentum, recorded in 2023). 
In Fig.\ref{fig:dYvsY} bottom left panel the same is shown, but for a typical 
high-multiplicity data set (Pb+Pb collisions at $150\,A\mathrm{GeV}/c$ 
beam momentum, recorded in 2022). Fig.\ref{fig:dYvsY} bottom right panel shows the 
same high-multiplicity data set with only a single sense wire switched on, thus 
eliminating the Cartesian combinatorial ghost hits using the design 
feature explained in Section~\ref{sec:grc}.

In order to further suppress the combinatorial background accompanying 
the correlation signal, the ROOT \cite{brun1997,rootsw} implementation of the 
Least Trimmed Squares (LTS) fitting was used. 
After calibrating MTPC-L, the other chambers in the TPC system are subsequently 
calibrated step-by-step against each other, using the already calibrated ones 
as geometric reference.

\begin{figure}[!ht]
  \begin{center}
    \includegraphics[height=8cm]{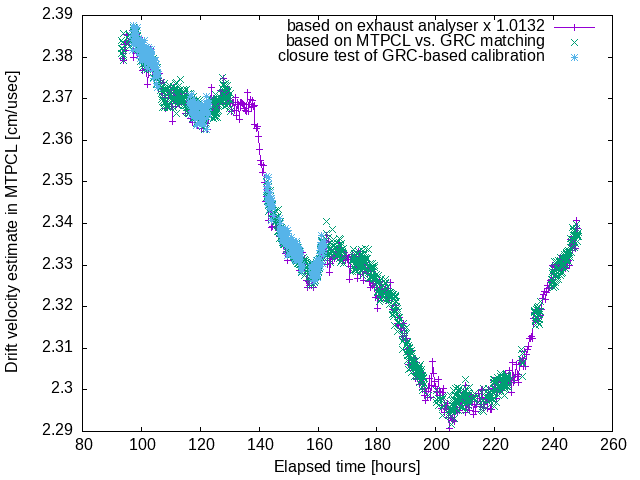}

    \vspace*{5mm}
    \includegraphics[height=8cm]{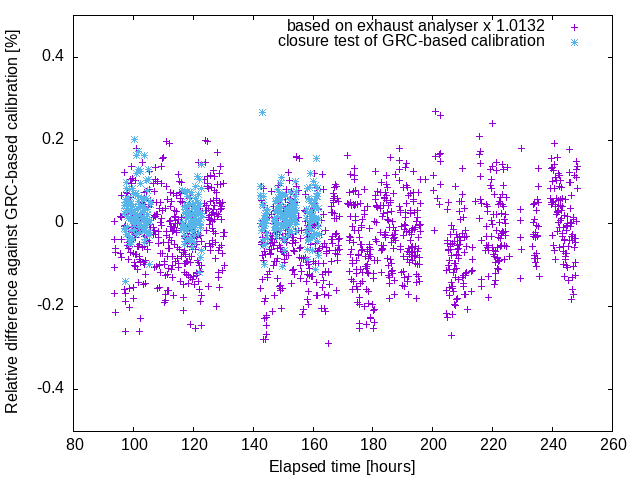}
  \end{center}
  \caption{(Color online) Top panel: final validation of the GRC-based 
    $\Delta{y}$ versus $y$ drift velocity calibration method. 
    Statistical uncertainties of the data points are below $0.05\%$, 
    therefore errorbars are not shown. It is seen that the closure test 
    is satisfied to the desired accuracy (1-2 permil). 
    That provides an empirical estimate for the systematic uncertainties. 
    Moreover, consistency is seen with the exhaust analyzer method, 
    up to a scale adjustment factor of $1.0132$. 
    Note that the exhaust analyzer method is expected to have some absolute 
    systematic error, which explains the necessity for the empirical scale 
    adjustment factor. 
    Bottom panel: the same data points, shown in terms of relative 
    difference with respect to the GRC-based calibration method.
  }
  \label{fig:performance}
\end{figure}

The statistical uncertainty of the method can be estimated from above 
by treating the pickup wire electrodes as $6\,\mathrm{mm}$ wide strips. 
Thus, the single-track position 
resolution of the GRC is not worse than $6\,\mathrm{mm}/\sqrt{12}$, corresponding to the resolution 
of a $6\,\mathrm{mm}$ wide uniform distribution. For a track sample of a ${\sim}5\,\mathrm{minute}$ 
time window, typically there are hundreds of MTPC-L tracks hitting the GRC, 
giving a resolution of $6\,\mathrm{mm}/\sqrt{12}/\sqrt{N_{\mathrm{tracks}}}$ 
for such ensemble of tracks. 
It is thus possible to push the statistical uncertainty below the desired 
$1\,\mathrm{mm}$, corresponding to $1$ per mil of the ${\sim}1\,\mathrm{m}$ 
TPC drift length. 
The systematic accuracy of the method was quantified using a closure test. 
By re-reconstructing the data using the calibrated drift velocities, the 
calculated correction on top of the previously-corrected drift velocities 
was found to be better than 1-2 permil. The systematics were also double-checked 
by comparing the obtained drift velocities to the exhaust analysis method, 
which agreed within 1-2 permil, up to a scale adjustment constant. Note that 
the exhaust analysis method is expected to have worse systematic uncertainties, 
hence the need for the scale adjustment constant for the exhaust method. 
These validation plots are seen in Fig.\ref{fig:performance}.

After calibrating the MTPC-L using the GRC, the other chambers are calibrated pairwise against each-other. 
Namely, the calibration scheme is MTPC-L from GRC, followed by VTPC-2 from MTPC-L, then 
MTPC-R from VTPC-2, and VTPC-1 from VTPC-2, and so on. 
Once having a methodology for drift velocity calibration, it is possible to extract further 
calibration constants from the data. Namely, it is possible to estimate the the 
trigger delay ($t_{0}$) and the drift direction displacement ($y_{0}$) of a chamber, 
see Appendix~\ref{sec:appT0Y0}. Moreover, using magnetic field-off calibration data, 
it is possible to estimate the chamber alignment parameters, as explained in 
Appendix~\ref{sec:appAlign}. A detailed systematic error analysis is also provided there. 
It is worth to note that after the above drift velocity and geometry calibration procedure, 
the magnetic field-off data are also used to map out the electric drift field distortions inside the TPCs: 
residuals of refitted local tracks are tabulated as a function of space, and are used as a 
fine-correction. At the boundaries, these residuals are up to ${\sim}1\,\mathrm{mm}$ and 
are much smaller inside the volumes. The drift field distortion corrections are of second order, and 
therefore do not interfere with the discussed drift velocity and geometry calibration.

\section{Concluding remarks}
\label{sec:conclusion}

In this paper a novel cost-efficient method was described for in-situ 
drift velocity monitoring in large volume TPCs. 
The method was deployed and is currently in use at the NA61/SHINE experiment at 
CERN. The key idea is to 
place a low-cost segmented detector, called to be the Geometry Reference 
Chamber (GRC), downstream of the TPC to be monitored. 
This monitoring solution was added retrospectively to the existing TPC 
system without a surgical operation to the chambers. A GRC 
was cost-effectively realized by using a MWPC designed to be compatible 
with existing TPC readout electronics. 
In such a way the development time and cost of the readout electronics 
could be spared. In order to match such an MWPC to the TPC readout electronics, 
it was necessary to slow down the signal formation in the GRC, 
since TPC readout electronics typically allows for a considerable trigger delay. 
This signal slowing 
was achieved by using an adjustable drift field transverse to the particle passage 
direction. The number of required readout channels was minimized using 
Cartesian readout. Since the NA61/SHINE experiment is used 
to study both low- and high-multiplicity collisions, the acceptance of the 
GRC was designed to be sufficiently large while addressing the issue of 
combinatorial ghost hits. The latter was handled 
by making the acceptance adjustable: when studying high-multiplicity collisions, 
individual sense wires can be switched off down to a single wire, eliminating the 
Cartesian ghosts, whereas for low-multiplicity runs the full acceptance is used 
with all of the sense wires operational.

The GRC system has been demonstrated of being capable of monitoring the 
drift velocity down to a one permil absolute precision in about $5$ minute 
time windows. That is, the pertinent method is applicable 
to large drift length TPC systems, with minimal design time, and 
relatively low manpower and cost.

Motivated by the GRC-based drift velocity monitoring method, a calibration 
procedure for the other geometric calibration parameters of the NA61/SHINE TPC 
chambers was also developed. 
Using that concept, all 8 parameters 
(time-dependent drift velocity, trigger delay, alignment shifts, and alignment angles) 
of the TPC system were successfully calibrated, for each chamber.

\section*{Acknowledgements}

We thank to the members of the REGARD gaseous R\&D group at the Wigner RCP, 
moreover for the support and cooperation from the members of the 
NA61/SHINE Collaboration at CERN, in particular to Bartosz Maksiak, 
Wojciech Bryli\'nski and Kyle Allison.

This work was supported by the Hungarian Scientific Research Fund 
(NKFIH OTKA-K138136-K138152, OTKA-FK135349 and TKP2021-NTKA-10).
Detector construction and testing was completed within the Vesztergombi 
Laboratory for High Energy Physics (VLAB) at HUN-REN Wigner RCP.

\appendix

\section*{Appendix}

\section{Trigger delay and drift-direction alignment of TPCs}
\label{sec:appT0Y0}

\begin{figure}[!ht]
  \begin{center}
    \includegraphics[height=6cm]{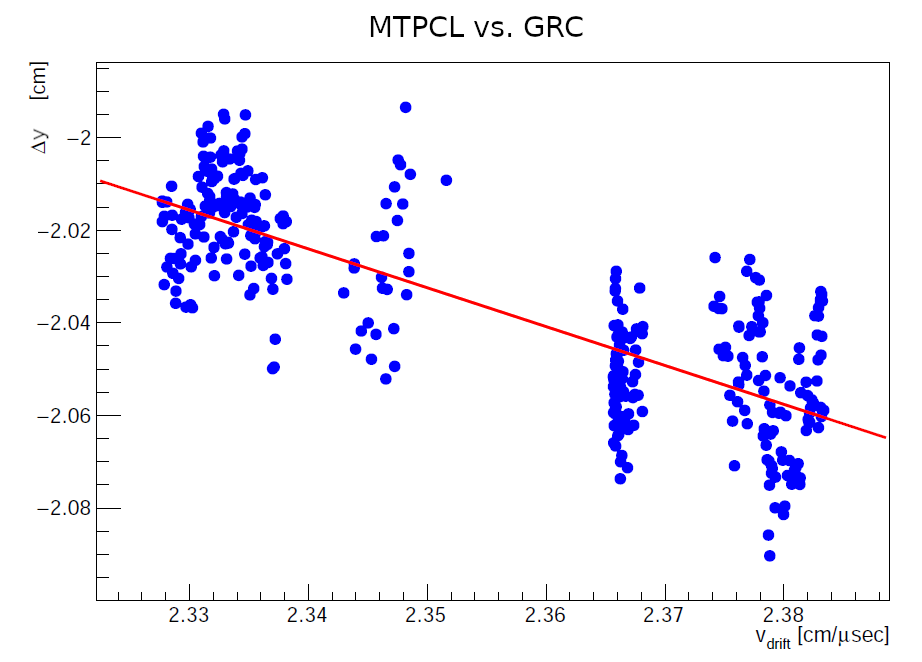}
  \end{center}
  \caption{(Color online) An example for the 
    $\Delta{y}$ versus $v_{\mathrm{drift}}$ analysis for trigger delay 
    ($t_{0}$) and drift direction shift ($y_{0}$) 
    calibration, after the drift velocity ($v_{\mathrm{drift}}$) has been 
    obtained. The slope of the pertinent scatter data gives the $t_{0}$ 
    correction, whereas the offset gives the $y_{0}$ correction. For the method 
    to be workable, data samples with different drift velocities (up to $2-3\%$) 
    are used. 
    The pertinent data set is a high multiplicity collision type 
    (Pb+Pb collisions at 
    $150\,A\mathrm{GeV}/c$ beam momentum, recorded in 2022).
  }
  \label{fig:dYvsVdrift}
\end{figure}

After the drift velocity has been calibrated with the $\Delta{y}$ versus 
$y$ method, one has $v_{\mathrm{drift,nom}}\approx v_{\mathrm{drift,true}}$, 
and therefore the TPC-GRC drift coordinate mismatch becomes $y$-independent, 
expressable as
\begin{eqnarray}
 \Delta{y} & = & \underbrace{(t_{0,\mathrm{true}}-t_{0,\mathrm{nom}})}_{\mathrm{slope}}\cdot v_{\mathrm{drift}} \;+\; \underbrace{(y_{\mathrm{anode,nom}}-y_{\mathrm{anode,true}})}_{\mathrm{offset}}
\label{eq:t0y0}
\end{eqnarray}
as seen from Eq.(\ref{eq:dYvsY}). Taking several data samples with actual 
drift velocity differing by $2-3\%$, the slope of the 
$\Delta{y}$ versus $v_{\mathrm{drift}}$ scatter plot provides an estimate 
for the additive 
correction between the true trigger delay ($t_{0,\mathrm{true}}$) 
and the trigger delay rough estimate applied during the initial reconstruction 
($t_{0,\mathrm{nom}}$). Thus, $t_{0,\mathrm{true}}$ can be estimated from 
the TPC-GRC mismatch data. After the $t_{0}$ is calibrated, the 
TPC-GRC mismatch will be not only $y$-independent, but also 
$v_{\mathrm{drift}}$-independent, and one is left with 
$\Delta{y}=y_{\mathrm{anode,nom}}-y_{\mathrm{anode,true}}$. Thus, 
the drift direction displacement ($y$-alignment correction) 
$y_{0}:=y_{\mathrm{anode,nom}}-y_{\mathrm{anode,true}}$ of 
the TPC versus the GRC can be estimated from the TPC-GRC mismatch data. 
An example for the $\Delta{y}$ versus $v_{\mathrm{drift}}$ calibration 
method is shown in Fig.\ref{fig:dYvsVdrift}.
Having the $t_{0}$ and $y_{0}$ obtained, the upstream TPC chambers can 
also be calibrated, using the already calibrated TPCs as geometry 
reference, in place of the GRC.

In the calibration database, for each TPC, the $t_{0}$ information is stored 
as follows. The $t_{0}$ for one of the TPCs (MTPC-L) is stored as a global 
reference (called to be the global $t_{0}$). For the other TPCs, only 
their relative $t_{0}$ difference with respect to this reference TPC is stored 
(called to be the chamber $t_{0}$). The rationale in such separation of the 
global and chamber $t_{0}$ contributions is the following. The chamber 
$t_{0}$ values are supposed to be detector constants, as they can only depend 
on cable lengths and electronic delays with respect to the trigger system. 
The global $t_{0}$, however, 
can also depend on the physics settings, such as beam time-of-flight, 
custom settings of delays inside the trigger logic etc. Thus, it is safer to 
recalibrate the global $t_{0}$ (MTPC-L versus GRC) for each data taking 
campaign, whereas it is enough to determine the relative chamber $t_{0}$ 
constants (other TPCs versus MTPC-L) only once, as they are detector constants. 
The recalculated relative chamber $t_{0}$ values are used, though, for 
consistency checks and quality monitoring.

\section{Systematic errors and alignment calibration of the TPCs}
\label{sec:appAlign}

As summarized in Section~\ref{sec:performance}, the final systematic errors 
of the in-situ drift velocity calibration via the $\Delta{y}$ versus $y$ method 
was verified to be not worse than 1-2 permil (or $1\,\mathrm{mm}$ overall, 
along the ${\sim}1\,\mathrm{m}$ drift length). This performance, however, 
was achievable after some detailed studies. 
Since the adjacent TPCs are calibrated against each-other 
(with first MTPC-L being calibrated against the GRC), 
the main source of systematic errors is the possible misalignment of the 
adjacent chambers (or the misalignment of MTPC-L versus the GRC). 
Considering a pair of adjacent chambers, one of which is an already fully 
calibrated chamber or the GRC, the other being an uncalibrated TPC, 
evaluation of affine transformations show that the track-by-track systematic 
error in the multiplicative drift velocity 
correction $v_{\mathrm{drift,corr}}:=v_{\mathrm{drift,true}}/v_{\mathrm{drift,nom}}$ 
caused by misalignment is
\begin{eqnarray}
 \mathrm{syst}(v_{\mathrm{drift,corr}}) = 
 \Big(y\cdot\theta_{x}+x\cdot\theta_{y}+(y-y_{\mathrm{mainvertex}})\cdot\theta_{x}+z_{0}\Big) \cdot \frac{1}{z-z_{\mathrm{mainvertex}}} \,,
\label{eq:syst}
\end{eqnarray}
to the first order.\footnote{This identity can be derived analytically, for magnetic 
field-off data, i.e.\ for straight tracks. Since the drift directions of 
the NA61/SHINE TPCs approximately coincide with the direction of the magnetic 
bending field, i.e.\ the drift and the bending does not interfere to the 
first order, the formula holds for magnetic field-on data as well, to a 
good approximation. In order to aid the tedious analytic evaluation of the three 
dimensional affine transformations, automatic formula manipulation software 
was used (the LinearAlgebra package of Maple).}
Here, $x,y,z$ is the particle hit position at the 
$z=const$ reference plane of the 
$\Delta{y}$ versus $y$ study (e.g.\ the GRC plane), moreover 
$x_{0},y_{0},z_{0}$ denote the (unknown) correction to the position alignment of the uncalibrated chamber, 
and $\theta_{x},\theta_{y},\theta_{z}$ denote the (unknown) correction to the angular alignment of the uncalibrated 
chamber, whereas the $x,y,z$ coordinates indexed by $()_{\mathrm{mainvertex}}$ 
denote the coordinates of the collision point. It is seen that 
the alignment caused drift velocity systematic error decreases with distance from the main-vertex, and therefore 
justifies our placement of the GRC downstream of all the TPCs, as seen in Fig.\ref{fig:DetectorSetup}. 
Even if the alignment of the MTPC-L with respect to the GRC were imperfectly known to 
$|y\cdot\theta_{x}|\lesssim 1\,\mathrm{cm}$ or $|x\cdot\theta_{y}|\lesssim 1\,\mathrm{cm}$ or 
$|(y-y_{\mathrm{mainvertex}})\cdot\theta_{x}|\lesssim 1\,\mathrm{cm}$ or 
$|z_{0}|\lesssim 1\,\mathrm{cm}$, this would not disturb 
the MTPC-L versus GRC drift velocity calibration beyond half permil, since 
$|z_{{}_{\mathrm{GRC}}}-z_{\mathrm{mainvertex}}|\approx 16\,\mathrm{m}$ is large. 
That is, the alignment imperfection does not give sizable systematic error to 
the MTPC-L versus GRC drift velocity calibration. 
When propagating the drift velocity calibration toward the upstream chambers 
(VTPC-1, VTPC-2), however, the systematic error Eq.(\ref{eq:syst}) by imperfectly 
known alignment can give a sizable contribution. Therefore, in this appendix 
we briefly describe, how the alignment is calibrated.

\begin{figure}[!ht]
  \begin{center}
    \includegraphics[height=4.4cm]{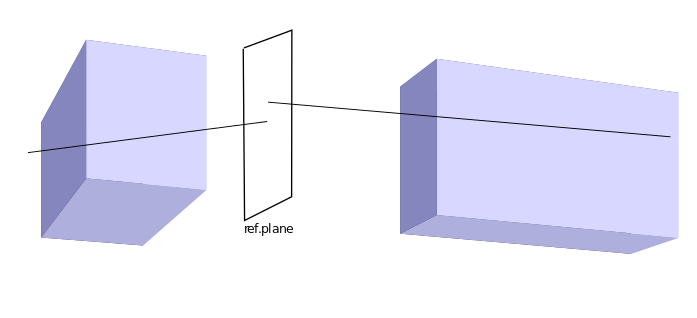}
  \end{center}
  \caption{(Color online) Sketch of the alignment calibration strategy for 
    adjacent TPCs: special magnetic field-off alignment calibration runs are recorded, 
    i.e.\ a tomography of the chamber system is performed by straight global main-vertex tracks. 
    If the relative alignment parameters for the chambers assumed during the reconstruction are imperfect, 
    the track parameters will have a mismatch, when compared on a joint reference 
    plane, for the adjacent chambers.
  }
  \label{fig:alignSketch}
\end{figure}

A TPC chamber has in total 8 unknown calibration coefficients related to its 
geometry: 
the $x_{0},y_{0},z_{0}$ position misalignment correction 
(transverse, drift direction, longitudinal), 
the $\theta_{x},\theta_{y},\theta_{z}$ angular misalignment correction, 
the drift velocity correction $(v_{\mathrm{drift,corr}}-1)$, 
and the correction to the trigger delay $t_{0,\mathrm{corr}}:=t_{0,\mathrm{true}}-t_{0,\mathrm{nom}}$. 
These unknown imperfection parameters, 
to be determined by the calibration procedure, are small, and therefore 
correction to the first order Taylor expansion terms with respect to these is enough. 
We use special magnetic field-off calibration runs in order to have only 
straight trajectories for charged particle tracks throughout the chamber system. 
In other words, a tomography of the chamber system is done using an ensemble of straight 
global tracks associated to the main-vertex, and their mismatch pattern is 
studied when dissected and refitted by straight lines in the local detectors, 
as illustrated in Fig.\ref{fig:alignSketch}.

\begin{figure}[!ht]
  \begin{center}
    \includegraphics[height=5cm]{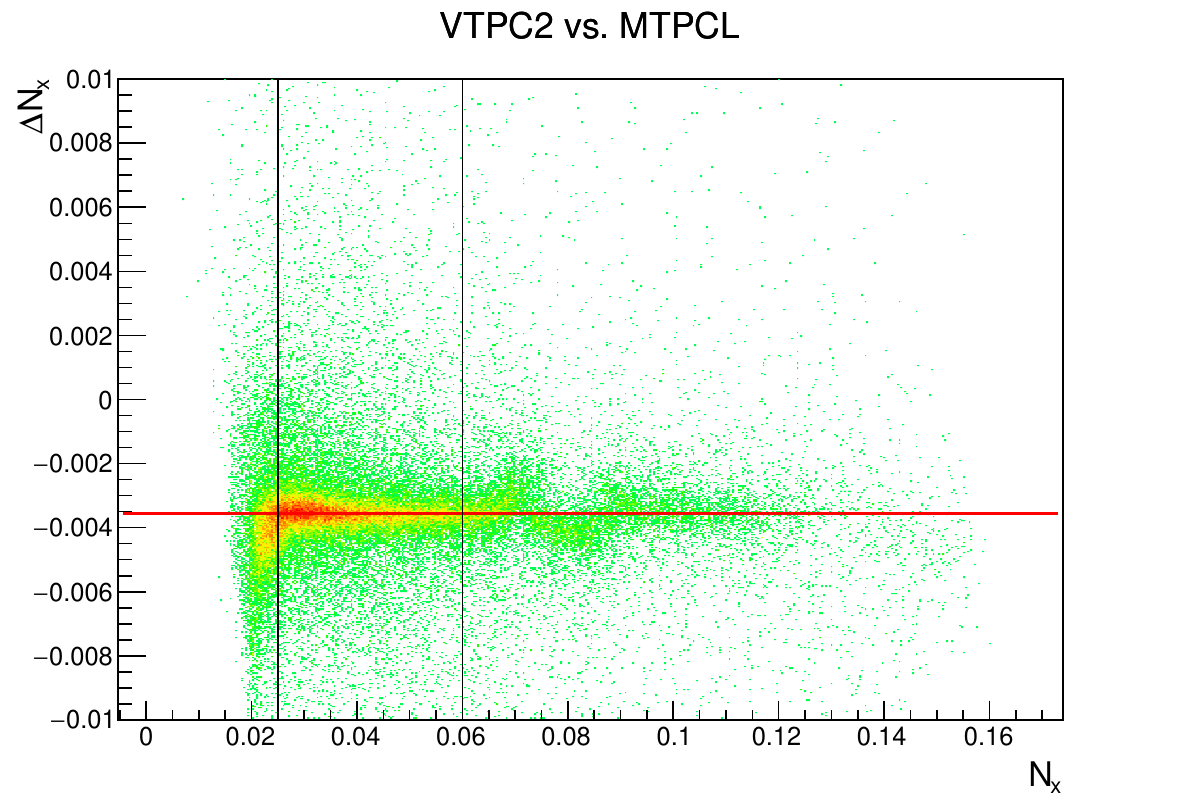}
    \includegraphics[height=5cm]{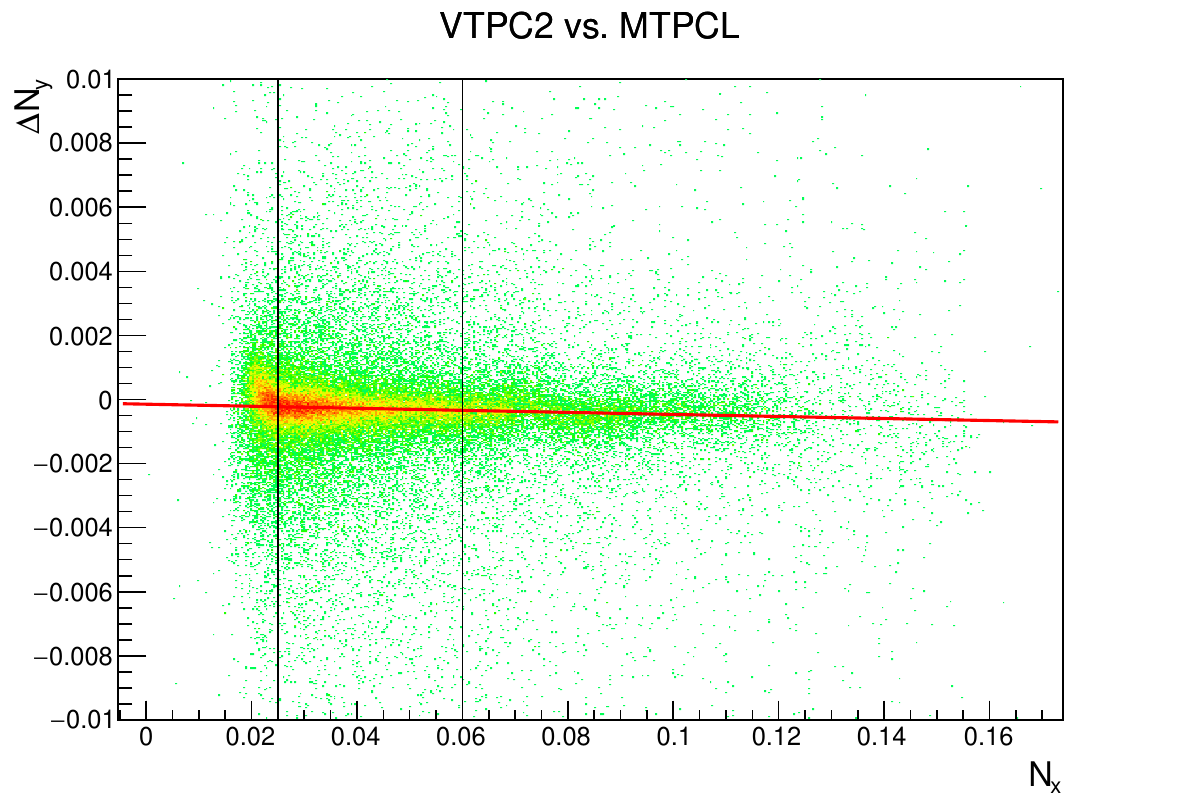}

    \vspace*{0.5cm}
    \includegraphics[height=5cm]{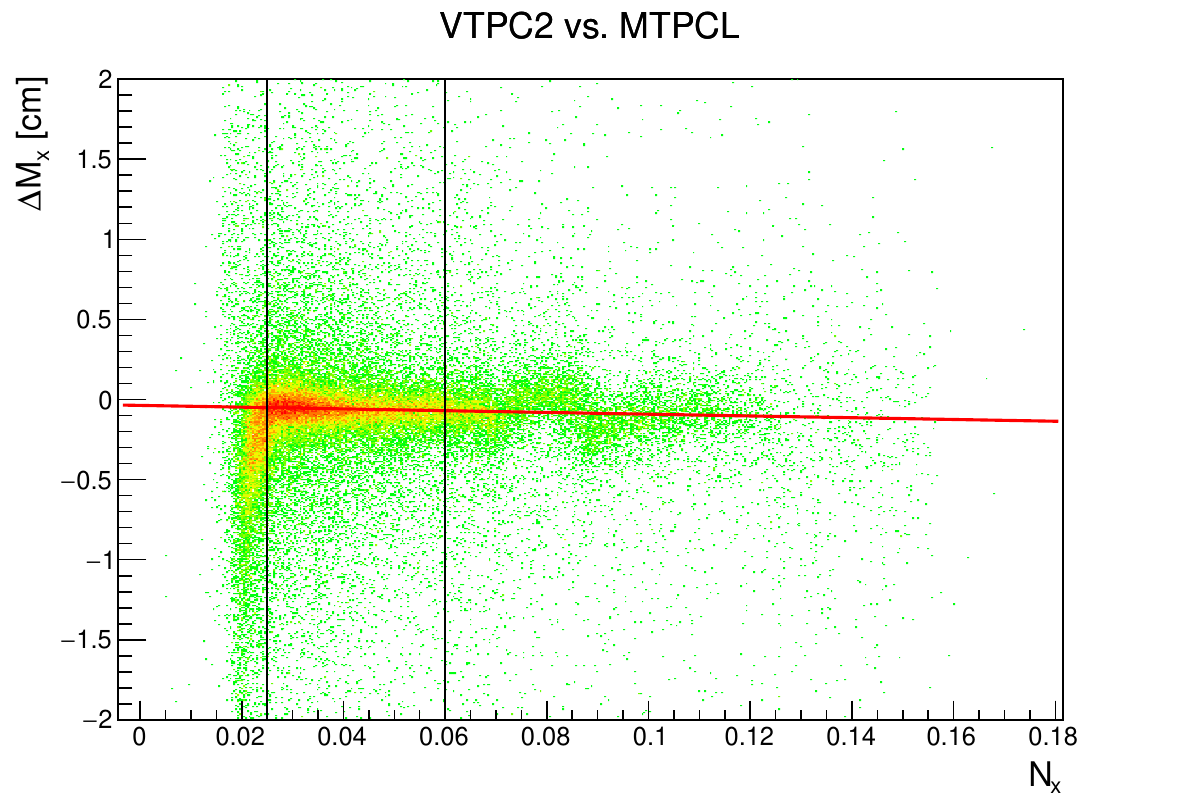}
  \end{center}
  \caption{(Color online) Top panels: an example for the 
    $\theta_{x},\theta_{y},\theta_{z}$ angular alignment calibration using 
    $\Delta{N}_{x}$ versus $N_{x}$ and $\Delta{N}_{y}$ versus $N_{x}$ plots 
    for $|N_{y}|\approx 0$ tracks in magnetic field-off data 
    (the pertinent sample is Pb+Pb at 
    $150\,A\mathrm{GeV}/c$ beam momentum, recorded in 2022). 
    Bottom panel: an example for the $x_{0},z_{0}$ position alignment calibration using 
    $\Delta{M}_{x}$ versus $N_{x}$ plot for tracks in magnetic field-off data 
    (using the same sample as for the top panels).
    The black vertical lines indicate fiducial cuts, in order to avoid regions 
    close to the TPC field cages, where the drift electric field can be imperfect 
    (inhomogeneous).
    The color scale of the scatter plots is merely for emphasizing hit density for the eye.
    The line on the plots corresponds to a straight line fit to the 
    scatter data, using the background tolerant LTS method.
  }
  \label{fig:align}
\end{figure}

Quantitatively, the magnetic field-off data based alignment calibration goes as follows. 
For MTPC-L, the $(v_{\mathrm{drift,corr}}-1)$ parameter is determined as in 
Section~\ref{sec:performance}, and the parameters 
$t_{0,\mathrm{corr}}$,$y_{0}$ are determined as in Appendix~\ref{sec:appT0Y0}, 
whereas $\theta_{x}=\theta_{y}=\theta_{z}=0$ and $x_{0}=z_{0}=0$ by 
convention, i.e.\ MTPC-L is taken as alignment reference. The other chambers 
are calibrated against each-other pairwise, for all the 8 parameters, assuming 
that one in the pair is already calibrated. In order to fix sign conventions, 
in this section it is assumed that the downstream chamber is calibrated, 
but analogous calibration equations hold when the upstream chamber is the calibrated one. 
An imperfection in the 
assumption about the 8 parameters of the upstream chamber causes a mismatch between locally refitted 
track pieces in the adjacent chambers, when compared on a $z=const$ 
reference plane, see again Fig.\ref{fig:alignSketch}. If the straight track segments are parameterized by the 
four parameters $N_{x},N_{y},M_{x},M_{y}$ as $x(z)=M_{x}+(z-z_{\mathrm{ref}})\cdot N_{x}$ 
and $y(z)=M_{y}+(z-z_{\mathrm{ref}})\cdot N_{y}$, then the locally refitted and compared 
local tracks will give a track parameter mismatch vector field 
$\Delta{N}_{x},\Delta{N}_{y},\Delta{M}_{x},\Delta{M}_{y}$ as a function of 
the track parameters $N_{x},N_{y},M_{x},M_{y}$ in the downstream chamber 
(which is already calibrated or reference). Evaluating affine transformations, one arrives at the formula
\begin{eqnarray}
 \Delta{N}_{x} & = & (1+N_{x}^{2})\cdot\theta_{y} + N_{y}\cdot(N_{x}\cdot\theta_{x}+\theta_{z}), \cr
 \Delta{N}_{y} & = & N_{y}^{2}\cdot(\theta_{x}+\theta_{y}) + N_{y}\cdot(v_{\mathrm{drift,corr}}-1) - N_{x}\cdot\theta_{z} + \theta_{x}, \cr
 \Delta{M}_{x} & = & N_{x}\cdot(M_{y}\cdot\theta_{x}+M_{x}\cdot\theta_{y}+z_{0}) + z_{\mathrm{ref}}\cdot\theta_{y} + M_{y}\cdot\theta_{z} - x_{0}, \cr
 \Delta{M}_{y} & = & N_{y}\cdot(M_{x}\cdot\theta_{y}+M_{y}\cdot\theta_{x}+z_{0}) + z_{\mathrm{ref}}\cdot\theta_{x} - M_{x}\cdot\theta_{z} - y_{0} \cr
               &   & \qquad + (M_{y}-y_{\mathrm{anode,nom}})\cdot(v_{\mathrm{drift,corr}}-1) - t_{0,\mathrm{corr}}\cdot v_{\mathrm{drift,nom}}
\label{eq:calib1}
\end{eqnarray}
for the mismatch field, to the first order, given the 8 imperfection parameters 
$\theta_{x},\theta_{y},\theta_{z}$, $x_{0},z_{0}$, 
$(v_{\mathrm{drift,corr}}-1)$, $t_{0,\mathrm{corr}}$, $y_{0}$. For practical 
reasons, the track sample satisfying $|N_{y}|\approx 0$, i.e.\ close to horizontal tracks 
are used, since for these the first two lines of Eq.(\ref{eq:calib1}) simplifies as
\begin{eqnarray}
 \mathrm{for\; }|N_{y}|\approx 0, \mathrm{one\; has}: & & \cr
 \Delta{N}_{x} & = & (1+N_{x}^{2})\cdot\theta_{y}, \cr
 \Delta{N}_{y} & = & - N_{x}\cdot\theta_{z} + \theta_{x}.
\label{eq:calib2}
\end{eqnarray}
That is, for magnetic field-off data, for $|N_{y}|\approx 0$ 
main-vertex tracks, from the $\Delta{N}_{x}$ versus $N_{x}$ plot the parameter 
$\theta_{y}$ can be read off, whereas from the 
$\Delta{N}_{y}$ versus $N_{x}$ plot the parameters $\theta_{x}$ and $\theta_{z}$ 
can be read off, see Fig.\ref{fig:align} top panels. After re-reconstructing with the obtained 
$\theta_{x},\theta_{y},\theta_{z}$ angular alignment corrections, the third line of 
Eq.(\ref{eq:calib1}) yields
\begin{eqnarray}
 \Delta{M}_{x} & = & N_{x}\cdot z_{0} - x_{0}.
\label{eq:calib3}
\end{eqnarray}
That is, for magnetic field-off data, from the $\Delta{M}_{x}$ versus $N_{x}$ 
plot the parameters $x_{0}$ and $z_{0}$ can be read off, 
see Fig.\ref{fig:align} bottom panel.

After re-reconstruction with the obtained $\theta_{x},\theta_{y},\theta_{z}$ 
angular alignment corrections and the $x_{0},z_{0}$ alignment shift corrections, 
the fourth line of Eq.(\ref{eq:calib1}) yields
\begin{eqnarray}
 \Delta{M}_{y} & = & M_{y}\cdot(v_{\mathrm{drift,corr}}-1) - y_{\mathrm{anode,nom}}\cdot(v_{\mathrm{drift,corr}}-1) - t_{0,\mathrm{corr}}\cdot v_{\mathrm{drift,nom}} - y_{0}
\label{eq:calib4}
\end{eqnarray}
which is nothing but the already known Eq.(\ref{eq:dYvsY}), to the first order in 
$(v_{\mathrm{drift,corr}}-1)$, re-expressed in the notations of this appendix. 
That is, the drift velocity correction 
$(v_{\mathrm{drift,corr}}-1)$ can be determined via the usual $\Delta{y}$ versus $y$ 
method, as already described in Section~\ref{sec:performance}. Finally, 
after re-reconstruction, one arrives at
\begin{eqnarray}
 \Delta{M}_{y} & = & - t_{0,\mathrm{corr}}\cdot v_{\mathrm{drift,nom}} - y_{0},
\label{eq:calib5}
\end{eqnarray}
which is nothing but the already known Eq.(\ref{eq:t0y0}), re-expressed in the notations of this 
appendix. Therefore, the remaining corrections $t_{0,\mathrm{corr}}$ and $y_{0}$ 
can be determined via the usual $\Delta{y}$ versus $v_{\mathrm{drift}}$ method, as already described in 
Appendix~\ref{sec:appT0Y0}. Thus, the triangular system of calibration equations 
for the 8 TPC geometry unknowns is saturated, and the entire TPC system can be calibrated.

After the full alignment correction, drift velocity correction, $t_{0}$ correction, 
$y_{0}$ correction, the overall systematic uncertainty of the geometry related calibration 
parameters of the large TPCs, as estimated via closure tests, are listed in Table~\ref{tab:syst}. 
The statistical uncertainties of the parameters are negligible.

\begin{table}[!ht]
  \begin{center}
    \begin{tabular}{r|ccccc}
             & $\theta_{x},\theta_{y},\theta_{z}$ [${}^{\circ}$] & $x_{0},z_{0}$ [$\mathrm{mm}$] & $v_{\mathrm{drift}}$ [$\%$] & $t_{0}$ [$\mathrm{ns}$] & $y_{0}$ [$\mathrm{mm}$] \\
             & syst. & syst. & syst. & syst. & syst. \\
  \hline
      MTPC-L & $0$ (reference) & $0$ (reference) & $0.2$ & $11$ & $0.15$ \\
      VTPC-2 & $0.038$         & $0.30$          & $0.2$ & $33$ & $0.25$ \\
      MTPC-R & $0.014$         & $0.20$          & $0.2$ & $33$ & $0.63$ \\
      VTPC-1 & $0.026$         & $0.25$          & $0.2$ & $32$ & $0.07$ \\
    \end{tabular}
  \end{center}
  \caption{The table of overall systematic uncertainties of the TPC geometry parameters, for 
    the large chambers, as estimated via closure tests. 
    The calibration scheme was GRC $\rightarrow$ MTPC-L, followed by MTPC-L $\rightarrow$ VTPC-2, then 
    VTPC-2 $\rightarrow$ MTPC-R and VTPC-2 $\rightarrow$ VTPC-1.
  }
  \label{tab:syst}
\end{table}

\end{document}